\documentclass{jetpl}
\twocolumn
\title{$z=3$ Lifshitz-Ho\v{r}ava model and Fermi-point scenario of emergent gravity}

\rtitle{$z=3$ Lifshitz-Horava model and Fermi-point scenario of emergent gravity}

\sodtitle{$z=3$ Lifshitz-Horava model and Fermi-point scenario of emergent gravity}

\author{G.E. Volovik
 $^{\#}$\/\thanks{
volovik@boojum.hut.fi} 
}

\rauthor{G.E.Volovik}

\sodauthor{Volovik}

\address{Low Temperature Laboratory, Helsinki University of
Technology, P.O.Box 5100, FIN-02015, HUT, Finland
\\
 Landau Institute for Theoretical Physics RAS, Kosygina 2,
119334 Moscow, Russia}

\dates{April 29, 2009}{*}

\abstract{Recently Ho\v{r}ava  proposed a model  for gravity which is described by the Einstein action in the infrared, but lacks the Lorentz invariance in the high-energy region where it experiences the anisotropic scaling.  We test this proposal using two condensed matter examples of emergent gravity: acoustic gravity 
and gravity emerging in the fermionic systems with Fermi points. We suggest that quantum hydrodynamics, which together with the quantum gravity is the non-renormalizable theory, may exhibit   the anisotropic scaling in agreement with the proposal. The Fermi point scenario of emergent general relativity demonstrates  that under general conditions,  the infrared Einstein action may be distorted, i.e.  the Ho\v{r}ava parameter $\lambda$ is not necessarily equal 1 even in the low energy limit.   The consistent theory requires special hierarchy of the ultra-violet energy scales and the fine-tuning mechanism for the Newton constant.   }

\begin{document}

\maketitle


\section{Introduction}

Recently Ho\v{r}ava  \cite{HoravaPRL2009,HoravaPRD2009,Horava2008} presented a candidate for quantum  gravity with anisotropy between space and time at high energy, and with emergent EinsteinÕs general relativity in the low-energy corner. It is assumed that the effective speed of light $c$, and the Newton constant $G$ both emerge from  the deeply nonrelativistic theory at short distances. At the moment, the Ho\v{r}ava model with anisotropic scaling is phenomenological. It does not
contain an explicit  mechanism of formation of general relativity in the infrared. Here we try to test the proposal using the condensed matter examples of emergent gravity, such as acoustic gravity emerging in hydrodynamics \cite{Unruh1981-1995}
and the Fermi point scenario, in which the Lorentz invariance emerges in the infrared together with the gauge fields and gravity \cite{FroggattNielsen1991,VolovikBook,HoravaPRL2005}.

\section{Particle spectrum}

The asymmetric scaling suggested by Ho\v{r}ava for the high-energy regime 
\cite{HoravaPRL2009,HoravaPRD2009}, implies that the spectrum of fermions or bosons extrapolates between 
the infrared macroscopic and ultra-violet microscopic behavior in the following way:
\begin{equation}
E^2(p)=p^2 + E_{\rm micro}^2(p) ~~,~~E_{\rm micro}(p)= \frac{p^z}{M^{z-1}}~.
\label{square_spectrum}
\end{equation}
Here we put  $c=1$ for the emergent `speed of light' -- the limiting velocity of the low-energy particles;  $M$  is  the mass parameter; and $z$ is the parameter of the anisotropic scaling at high energy. In  Ref. \cite{Horava2008} the anisotropic scaling with  $z=2$ has been used, while  in Refs. \cite{HoravaPRL2009,HoravaPRD2009} --  that with $z=3$. \footnote{
The spectrum $E_{\rm micro}(p)$ with $z=2$ occurs at the Lifshitz point in the theory of phase transitions  \cite{Lifshitz1941}. This is  the point on the phase diagram,
where the parameters $a_0$ and $a_1$ of the expansion of the spectrum 
$\omega^2(k)= a_0 +a_1 k^2 + a_2 k^4$ vanish, $a_0=a_1=0$. Near this point the dominating term in the  spectrum is $\omega(k)\approx  a_2^{1/2} k^2$. This the reason why the Ho\v{r}ava model anisotropic scaling  is called the Ho\v{r}ava-Lifshitz gravity. Cosmology with Lifshitz scalar field obeying the $z=3$ anisotropic scaling at high energy has been discussed in Ref. \cite{Calcagni2009}.
}

Below, from the estimation of the Newton constant $G$ in effective gravity, we shall show that $M$ must be somewhere in between the GUT and the Planck energy scales: $E_{\rm Planck}> M > E_{\rm GUT}$.
We need also the ultra-violet energy cut-off, at which Eq. (\ref{square_spectrum}) is violated at even higher energies. By definition, this ultra-violet  energy scale must be much  larger than $M$, i.e.
$E_{\rm uv}\gg M$. Theories  with $E_{\rm uv}\gg E_{\rm Planck}$ can be found e.g. in Refs. \cite{KlinkhamerVolovik2005} and \cite{Laperashvili2003}.  
For the  $z=3$ anisotropic scaling, the spectrum at large $p$ is cubic:
\begin{equation}
E(p)\approx E_{\rm micro}(p) = \frac{p^3}{M^{2}}~~,~~E_{\rm uv}\gg p\gg M~.
\label{spectrum_large_p}
\end{equation}

In the Ho\v{r}ava model it is assumed that Lorentz invariance emerges at low $p$, and the spectrum in Eq. (\ref{square_spectrum}) becomes linear:
\begin{equation}
E(p) \approx E_{\rm macro}(p) = p~~,~~ p\ll M~.
\label{spectrum_small_p}
\end{equation}
In the Fermi point scenario, the Lorentz invariance is not an accidental
symmetry of a low-energy theory. It naturally emerges at low energy because it is dictated by the topologically stable Fermi point  in the fermionic spectrum. In the low energy corner the momentum approaches the Fermi point at $p=0$, and in the vicinity of the Fermi point the spectrum is necessarily linear  
\cite{FroggattNielsen1991,VolovikBook,HoravaPRL2005,Creutz2008}. Gravity and gauge fields are also not accidental in the Fermi point scenario: they naturally emerge as the low energy collective modes interacting with the low energy relativistic fermions living near the Fermi point. 

 \section{Lorentz symmetry violation}

As follows from Eq.(\ref{spectrum_small_p}), the energy $M$ is the scale at which the Lorentz invariance is violated. As we shall see below,  the estimate of the Newton constant suggests that the Lorentz symmetry breaking scale $M$ must be below or on the order of the Planck scale. Fortunately, for the $z=3$ scaling this does not contradict to observations. This is because for $z=3$, the Lorentz violating  corrections to the linear spectrum at small $p$ 
\begin{equation}
E(p)=p\sqrt{1+\frac{p^4}{M^4}} \approx   p\left(1+\frac{p^4}{2M^4}+\ldots\right) ~~,~~ p\ll M~,
\label{corr_lin_spectrum}
\end{equation}
start  with the $p^5/M^4$ term. In the  typical Lorentz violating models, including the Ho\v{r}ava $z=2$ scaling \cite{Horava2008}, the  Lorentz violation starts with the natural $p^3/M^2$ term: $E=p(1 + p^2/M^2 +\ldots)$, see e.g. \cite{Jannes2009}. The $p^3/M^2$ term in spectrum  with $M$ of order of Planck scale and with the prefactor of order  unity is strongly forbidden by observations: if  $M$ is of order of Planck scale or smaller, the prefactor must be smaller than $10^{-6}$ (see e.g. \cite{MaccioneLiberati2008}). But the $p^5/M^4$ term is acceptable, which is one of the  advantages of the $z=3$ scaling.

The other advantage of the $z=3$ scaling is the  convergence of the integrals over momentum $p$  in the ultra-violet limit. This is in particular because at large $p$ the corrections to the cubic spectrum rapidly decay:
\begin{equation}
E(p)  = E_{\rm micro}\sqrt{1+\frac{M^4}{p^4}} \approx E_{\rm micro} \left(1+\frac{M^4}{2p^4}+\ldots\right)~~,~~ p\gg M~.
\label{corr_lin_spectrum}
\end{equation}
This makes the $z=3$ model  potentially ultra-violet  complete.

\section{Anisotropic scaling in  quantum hydrodynamics}

The first quantization scheme for hydrodynamics was suggested by Landau in 1941
when he developed the theory of superfluidity in liquid $^4$He \cite{Landau1941}.
However, later it appeared that beyond the quantization of sound waves, quantum hydrodynamics has the same  ultra-violet problems as quantum gravity and is also non-renormalizable. 
It was suggested that in  quantum hydrodynamics, the corrections to the linear spectrum  of sound waves may also start with the $k^5$ term:  \cite{Volovik2008}
\begin{equation}
 \omega(k)=c_sk + \gamma \frac{\hbar}{\rho}k^5 + \ldots~.
\label{phonon_spectrum}
\end{equation}
Here $c_s$ is the speed of sound,  $\rho$ is the mass density of a liquid and $\gamma$ is the factor of order unity. 

The important property of this equation is that the quantum correction does not contain the `speed of light' $c_s$ and is solely determined by the quantities which enter the commutation relations in quantum hydrodynamics: Planck constant $\hbar$ and mass density $\rho$ (see Ref.  \cite{Landau1941}).
This may give us a guiding principle for the spectrum in quantum hydrodynamics  at high frequency:  in the high-frequency regime the vortex-like degrees of freedom are dominating (whose quantum dynamics has been discussed in Ref.  \cite{RasettiRegge1975}). The acoustic degrees of freedom are not relevant, the speed of sound $c_s$ drops out of equations and thus only   $\rho$ and 
$\hbar$ enter the spectrum. Based on this principle, it is possible to construct anisotropic scaling with different $z$. If one uses the dimensionality analysis  only, then the  high frequency spectrum which  depends on $\rho$ and  $\hbar$ and does not depend on $c_s$ must be like that: 
\begin{equation}
 \omega_{\rm micro}(k)= \gamma \frac{\hbar}{\rho}k^5 ~~,~~k\gg \left( \frac{\rho c_s}{\hbar}\right)^{1/4}~.
\label{phonon_spectrum_micro}
\end{equation}
This matches  Eq.(\ref{phonon_spectrum}) at low $\omega$, but  corresponds to the anisotropic scaling with $z=5$, rather than that with $z=3$.

Alternatively,  one may take into account the acoustic metric for phonons propagating 
in the liquid \cite{Unruh1981-1995}. For the liquid at rest one has:
 \begin{equation}
 g^{\mu\nu}_{\rm acoustic}=
 \mathrm{diag}\left(-\frac{1}{\rho c_s},\frac{c_s}{\rho},\frac{c_s}{\rho},\frac{c_s}{\rho}\right)
 \,,
\label{eq:acoustic}
\end{equation}
\begin{equation}
g_{\mu\nu}^{\rm acoustic}=
 \mathrm{diag}\left( - \rho c_s,\frac{\rho}{c_s},\frac{\rho}{c_s},\frac{\rho}{c_s}\right)
 \,,
\label{eq:acoustic_cov}
\end{equation}
Then one can write the following equation for spectrum:
 \begin{equation}
 g^{\mu\nu}_{\rm acoustic}k_\mu k_\nu + \frac{2\gamma \hbar}{\rho^2} k^6 = 0
 \,.
\label{eq:spectrum_acoustic}
\end{equation}
This equation has the correct  covariant form at low frequency; does not contain $c_s$ explicitly (it is hidden in the metric); and matches  the expansion in Eq.(\ref{phonon_spectrum}). But now it corresponds to the anisotropic scaling with  $z=3$ suggested by Ho\v{r}ava for quantum gravity.

All this is highly speculative, since the full quantization of classical hydrodynamics has not yet been constructed. That is why it is not clear whether it is reasonable to  apply the above guiding principle to quantum gravity. One may suggest for example that in the high-frequency regime the emergent speed of light $c$ is dropped out of equations and the spectrum may depend only on the energy density of the vacuum $\epsilon$. Then the  dimensionality analysis  gives the inverse cubic spectrum for large momentum,  $E_{\rm micro}(p) \propto   \epsilon \hbar^3 p^{-3}$. This might be reasonable, since it is consistent with the estimate of the energy density of the vacuum: $\epsilon \propto \int d^3p/(2\pi \hbar)^3~E_{\rm micro}(p)$.

\section{Newton constant in Fermi point scenario}

In the Fermi point scenario, the 3+1 gravity emerges at small $p$  together with gauge fields \cite{VolovikBook,HoravaPRL2005} (the 2+1 gravity emerging in the vicinity of  the Dirac point in graphene is discussed e.g. in Ref. \cite{Huan2007}).  The action for effective gravity and effective gauge fields is obtained by integration over fermions in the same manner as in Sakharov induced gravity \cite{Sakharov} and Zeldovich induced gauge field \cite{Zeldovich1967b}. Let us start with gravity and show that the scale $M$ in the spectrum (\ref{square_spectrum}) for the $z=3$ anisotropic scaling is on the order of or even below the Planck energy scale $E_{\rm Planck}$.

The gravitational coupling can be estimated using  the Sakharov approach in which the action for gravity is obtained by  integration over  quantum fluctuations   \cite{Sakharov}.  The natural
regularization in the Sakharov equation for the Newton constant,   $1/G \sim \int \frac {d^3 p}{p}$, 
leads to $1/G \sim M^2 \ln (E_{\rm uv}^2/M^2)$. 
We illustrate this using two different ways of regularization.
In the first illustration, we substitute $p$ in denominator by $E(p)$ in Eq.(\ref{square_spectrum}). For the $z=3$ anisotropic scaling this gives 
\begin{equation}
 \int \frac {d^3 p}{p} \rightarrow G_{\rm ren}^{-1}  \sim \int \frac{d^3p}{E(p)}  
 \sim M^2 \ln \frac{E_{\rm uv}^2}{M^2}~~.
\label{G3}
\end{equation}

In the second case  we substitute $1/p$ by $E(p)/p^2$. Now the inverse Newton constant diverges at high energy as $E_{\rm uv}^4/M^2$. But  gravity exists only at low energy, and thus in the same manner as in the Casimir effect one may subtract the diverging ultra-violet contribution coming from the $E_{\rm uv}$ scale.  Then for $z=3$ one obtains the same estimate as in Eq. (\ref{G3}):
\begin{eqnarray}
 \int \frac {d^3 p}{p}\rightarrow  G_{\rm ren}^{-1}  \sim \int \frac{d^3p}{p^2} \left(E(p)-E_{\rm micro} (p)  \right) \sim
\label{G1}
\\
 \sim \frac{1}{M^2}\int  d^3p ~p \left(\sqrt{ 1 + \frac{M^4}{p^4}}-1 \right)
 \sim M^2 \ln \frac{E_{\rm uv}^2}{M^2}~~.
\label{G2}
\end{eqnarray}

The same consequence of two different regularization schemes suggests that the estimate $1/G \sim M^2 \ln (E_{\rm uv}^2/M^2)$ for the $z=3$ anisotropic scaling remains valid in direct calculations of $G$.  Taking into account the large logarithm in Eq. (\ref{G3}) and the large number of fermionic species in the Standard Model, one obtains that the Lorentz violating  scale $M$ is somewhere between the Planck and GUT energy scales, $E_{\rm GUT}<M<E_{\rm uv}$.

In both regularization schemes, the integral is concentrated at the energies $E$ on the order of $M$ and higher. At $E\sim M$, the fermions are not Lorentz invariant. The part of action for the induced gravity, which is obtained by integration over the non-relativistic  fermions, is  non-covariant  even in the low energy limit. This means  that  gravity induced in the low-energy corner does  not obey the Einstein equations:  the parameter  $\lambda$ in Ho\v{r}ava action is not equal unity even in the infrared; the limiting velocity of the low frequency gravitons $c_g$ does not coincide with  the fermionic  `speed of light'  $c$;
\footnote{
In the Ho\v{r}ava approach, gravity  is universal and thus the metric $g_{\mu\nu}$ and the speed of light $c$ are the same for all species of matter. However, in the effective theory it is not  guaranteed that in the infrared the graviton  limiting velocity $c_g$ coincides with $c$.
} 
gravitons acquire mass of order $M$; and other unpleasant things. 

Note that in the anisotropic $z=3$ scaling, the vacuum energy diverges more strongly than for the isotropic $z=1$ scaling:  $\epsilon \propto \int d^3p E(p) \sim E_{\rm uv}^6/M^2$. However, this divergence is not crucial since  in the perfect equilibrium vacuum the contribution of the zero point energy of quantum fields  to the cosmological constant will be fully compensated by the contribution from the microscopic (ultra-violet) degrees of freedom above $E_{\rm uv}$ \cite{VolovikBook,KlinkhamerVolovik2008b}. This follows purely from the thermodynamic analysis of the equilibrium Minkowski vacuum and does not depend on the details, such as index $z$ and parameters $M$ and $E_{\rm uv}$ (see also Ref. \cite{Thomas2009} and references therein). That is why the issue of the cosmological constant problem \cite{Nastase2009,Takahashi:2009wc,Sotiriou2009} is irrelevant for the model. It is the Newton constant which requires the fine-tuning (see below), rather than the cosmological constant for which the fine-tuning occurs automatically in any self-sustained vacuum \cite{KlinkhamerVolovik2008b}. 

\section{Discussion}

The same problem with the non-invariant terms in the low-energy action arises in the effective action for the induced gauge field, which is also obtained by integration over the fermionic fields. Following the Zeldovich prescription \cite{Zeldovich1967b}, one obtains that with the logarithmic accuracy the running coupling calculated at the energy scale of the order of $Z$-boson mass is $\alpha^{-1}(M_{Z})\sim   \ln (M^2/M_Z^2)$.
This contribution to $\alpha^{-1}$ comes from integration over fermions in the region of energies $M_Z<E<M$. The main logarithmic term in the action is Lorentz invariant due to the contribution from the region $M_Z \ll E \ll M$, where fermions are relativistic. However,  the non-logarithmic terms in the action which come from the region $E\sim M$ are not invariant. The relative magnitude of the Lorentz violating terms in the effective electrodynamics is thus on the order of $ 1/   \ln (M^2/M_Z^2) \sim \alpha$. This highly contradicts to observations: the relative magnitude of the Lorentz violating terms in electrodynamics is experimentally smaller than $10^{-15}$ (see Ref. \cite{Klinkhamer2008}).

That is why one needs a mechanism which protects against the non-covariant terms in the action for gravity and gauge fields emerging in the infared.
The smallness  of the non-covariant terms  can be achieved  if there is some small parameter in the theory \cite{Bjorken2001}. For example, if the Lorentz violating scale is much higher than the Planck scale, $E_{\rm LV} \gg E_{\rm Planck}$,  the small parameter $E_{\rm Planck}^2/E_{\rm LV}^2$ will suppress the Lorentz violating terms (see Ref.  \cite{KlinkhamerVolovik2005}). This suggests that the Lorentz violating scale $M$ in the $z=3$ Ho\v{r}ava model must be very much larger than the Planck scale, $M \gg E_{\rm Planck}$. But this contradicts to the Eq.(\ref{G3}) for the Newton constant, which says that   $M$ is below the Planck scale, $M<E_{\rm Planck}$. If one insists that $M \gg E_{\rm Planck}$,  one needs in an additional principle or symmetry in the underlying microscopic physics, which may reduce the inverse Newton constant $1/G$ from its natural value   $M^2$ in Eq.(\ref{G3}) to its experimental value  $E_{\rm Planck}^2$.

In conclusion, there is a hint from quantum hydrodynamics on the possibility of the $z=3$ anisotropic scaling in acoustic gravity. However, to support the Ho\v{r}ava proposal one must find the proper mechanism, by which the gauge fields, gravity and relativistic fermions emerge together at the low energy fixed point. The Fermi point scenario of emergent relativity may provide such a mechanism. But the problem within this scenario is that under general conditions  the Einstein action is  distorted even in  the low-energy corner. To restore the general relativity in the infrared,  one must construct the consistent hierarchy of the energy  scales ($M$,  $E_{\rm Planck}$, $E_{\rm LV}$ and $E_{\rm uv}$) and find the physically motivated mechanism for the fine-tuning of the Newton constant $G$. 

It is a pleasure to thank Frans Klinkhamer and Alexei Starobinsky for useful comments. This work is supported in part by the Russian Foundation
for Basic Research (grant 06--02--16002--a) and the
Khalatnikov--Starobinsky leading scientific school (grant
4899.2008.2).


\end{document}